\documentstyle[preprint,prl,aps]{revtex}

\begin{document}

\draft

\title{Neutrino-nucleus reactions on $^{12}$C and $^{16}$O}

\author{N. Auerbach$^1$,  N. Van Giai$^2$ and O.K. Vorov$^1$}  

\address{$^1$ School of Physics and Astronomy, Tel Aviv University, 
        Tel Aviv 69978, Israel}

\address{$^2$ Division de Physique Th\'eorique, Institut de Physique 
Nucl\'eaire, 91406 Orsay Cedex, France}

\date{May 1, 1997}

\maketitle

\begin{abstract}
Exclusive and inclusive $(\nu_\mu, \mu^-)$,  $(\nu_e, e^-)$ cross-sections 
and $\mu^-$-capture rates are calculated for $^{12}$C and $^{16}$O 
using the consistent random phase approximation (RPA) and pairing 
model. After a pairing correction is introduced to the RPA results 
the flux-averaged 
theoretical $(\nu_\mu, \mu^-)$,  $(\nu_e, e^-)$ cross-sections 
and $\mu^-$-capture rates in $^{12}$C are in good agreement 
with experiment. In 
particular when one takes into account the experimental error bars, the 
recently measured range of values for the $(\nu_\mu, \mu^-)$ cross-section 
is in agreement with the present theoretical results. 
Predictions of $(\nu_\mu, \mu^-)$ and  $(\nu_e, e^-)$ cross-sections 
in $^{16}$O are also presented.
\end{abstract}

\pacs{23.40.Bw, 25.30.Pt, 21.60.Jz}

\narrowtext

  Investigations of neutrino-nucleus interactions are presently in the
forefront of particle and nuclear astrophysics research. The study of
basic properties of the neutrino such as its mass, possibility of 
decay or oscillations, is probed using the interaction of neutrinos
with nuclei. The neutrino interaction with matter is very weak and therefore
neutrinos that are emitted from the inner parts of a star
are ``messengers" that bring information about the processes occurring 
inside the stars\cite{bah}. 
These extraterrestrial neutrinos are detected in reactions involving
nuclear targets. It is of considerable importance to provide a theoretical
framework that takes into account the various aspects of nuclear structure
and is able to describe well these reactions . Attempts of this kind were made
in the past\cite{wal} and in the last few  years such attempts have been 
revived\cite{kub,kol} in view of the availability of new 
experimental results in this
field of  neutrino-nucleus interaction. In particular we will refer 
here to the KARMEN results and the results obtained at the Liquid 
Scintillator Neutrino Detector (LSND) in Los Alamos. The 
$^{12}$C($\nu_{\mu}, \mu^-$)X measurements at the LSND were made 
with the same experimental 
setup as the one used in the neutrino oscillation experiment - the results 
of which were recently published\cite{ath96}. A study of the $^{12}$C(
$\nu_{\mu}, \mu^-$)X reactions
will shed indirectly some light on the question of validity of the
neutrino oscillation experiments. For example a strong disagreement between
theory and the experimental results for the $^{12}$C($\nu_{\mu}, \mu^-$)X 
could probably 
lead to some skepticism 
concerning the experimental setup of the neutrino 
experiments in general. 
In fact, the recently published theoretical results\cite{kol} for the 
inclusive $^{12}$C($\nu_{\mu}, \mu^-$)X are in disagreement, 
by a factor of two, 
with the LSND result\cite{alb}. 

\noindent
In the present paper we examine a range of neutrino
-nucleus cross-sections as well as $\mu^-$-capture rates 
with a special emphasis on the$^{12}$C case. Our calculations are 
performed  
first in the Hartree-Fock (HF) approximation and then in the consistent 
Hartree-Fock-
Random Phase Approximation (HF-RPA). The consistency refers here to the fact
that the HF mean field and the particle-hole (p-h) interaction result from the
same effective nucleon-nucleon, two-body force. 
Such RPA calculation preserves 
the energy-weighted sum rule of the strength distribution of one-body
operators\cite{boh} and therefore it is a favorable scheme when it 
comes to calculate 
the distribution of total strength. One is led to expect that, 
for closed shell nuclei inclusive 
cross-sections of processes governed by one-body transition operators
should be well reproduced in an RPA-type calculation. 
Indeed, the 
RPA had much success in providing an adequate description of giant
resonances\cite{ber,mek} and of a variety of inclusive processes 
in nuclei with good closed shells. 
An example relevant to our subject of neutrino-nucleus interactions is the 
inclusive $\mu^-$-capture on nuclei. 
The HF-RPA was able to reproduce\cite{mek} the inclusive capture
rates in a number of closed shell nuclei. 
It was found that the collectivity and RPA 
ground state 
correlations are very important in reaching good agreement with experiment. 
This suggests that other inclusive neutrino 
processes such as for example the ($\nu_{\mu},\mu^-$) reaction 
(which is the inverse to $\mu$-capture) will be well described by the 
HF-RPA.    
It is known however that the ground state of $^{12}$C is not a 
good closed shell. Admixtures of the $(p_{1/2})^2 (p_{3/2})^2$ 
configuration into the $(p_{3/2})^4$ configuration are large and one 
should expect substantial corrections to the matrix elements obtained 
in the RPA. As we will see these corrections play a very important 
role when one calculates the $(\nu_l, l^-)$ cross-sections, in 
particular the exclusive ones to the ground state of the daughter 
nucleus. The inclusive cross-sections to the excited states are 
less affected but, 
in order to achieve agreement with 
experiment one must 
nevertheless include these corrections also for the excited 
states. 

The numerical applications are performed with the Skyrme forces 
SGII and  SIII\cite{bei75,gia81}. 
The interaction SGII was adjusted so as 
to give the correct value of the Landau parameter $F'_0$\cite{gia81} in the 
spin-isospin particle-hole channel which should be of particular 
importance in charged current neutrino reactions on nuclei. 
First, the Hartree-Fock equations are solved in coordinate space to 
obtain the self-consistent mean field. This mean field determines the 
single-particle spectrum. For the 
present problem it is not necessary to treat exactly the 
single-particle continuum since we are not studying specific 
(exclusive) channels where a nucleon would be emitted following the 
($\nu_{\mu},\mu^-$) or ($\nu_{e},e^-$) reaction. Therefore, it is 
convenient to discretize the single-particle spectrum by diagonalizing 
the Hartree-Fock mean field on a harmonic oscillator basis. 
The reference Hartree-Fock state $\vert HF \rangle$ corresponds to the 
target nucleus, e.g., $^{12}$C. 
Let us denote by $i,j$,... ($a,b$,...) the proton (neutron) occupied 
states and by $I,J$,... ($A,B$,...) the proton (neutron) unoccupied 
states. The proton (neutron) creation and annihilation operators are 
respectively $p^+_i$ and $p_i$ ($n^+_i$ and $n_i$). In reactions of 
the ($\nu_{\mu},\mu^-$) or ($\nu_{e},e^-$) type the final states 
$\vert \lambda \rangle$  
belong to the $\Delta T_Z$= -1 daughter nucleus (e.g., $^{12}$N) and 
they can be described by the charge-exchange RPA\cite{aue83,col} model:
\begin{eqnarray}
\vert \lambda \rangle & = & \Big(\sum_{I,a} X^{(\lambda)}_{Ia}~p^+_I n_a 
+ \sum_{i,A} Y^{(\lambda)}_{iA}~p^+_i n_A \Big) \vert \tilde 0\rangle~,
\label{eq2}
\end{eqnarray}
where $\vert \tilde 0\rangle$ is the correlated RPA ground state. 
The $X^{(\lambda)}$ and $Y^{(\lambda)}$ are solutions of the charge-exchange 
RPA equations~\cite{aue83,col}.
For a one-body charge-exchange operator of the general form:
\begin{eqnarray}
 O & = & \sum_{\alpha,\beta} O_{\alpha \beta} p^{+}_\alpha n_{\beta}~,
\label{eq6}
\end{eqnarray}
the transition amplitude $\langle \lambda \vert F \vert \tilde 0 
\rangle$ can be expressed simply as:
\begin{eqnarray}
\langle \lambda \vert O \vert \tilde 0 \rangle & = & \sum_{Ia} 
X^{(\lambda)*}_{Ia} O_{Ia} - \sum_{iA} 
Y^{(\lambda)*}_{iA} O_{iA}~. 
\label{eq7}
\end{eqnarray}

In the case of a parent nucleus with zero angular momentum in 
the ground state the cross-section is\cite{wal,kub}:
\begin{equation}
\label{CROSS-SECTION}
\sigma = 
\frac{G^2}{2 \pi} cos^2 (\theta_C) \sum_{\lambda}
p_l E_l {\cal F}(Z,E_l) \int_{-1}^{1} d(cos (\theta)) 
{\cal M}_{\lambda 0}~,
\end{equation}
where $G$ and $\theta_C$ are the Fermi constant and the Cabibbo angle,
$p_l$ and $E_l$ are the momenta and energies of outgoing leptons
(muon or electron), $\theta$ is 
the angle between the momenta of the 
lepton and the incoming neutrino. The factor ${\cal F}$ accounts
for the effects of the final state interaction (FSI) of the
outgoing lepton with the daughter nucleus of charge $Z$\cite{wal,kub}. 
For the case of $(\nu_e, e^-)$ reactions 
the mass of the outgoing lepton is small 
and the effect is not so important.
The effect of the FSI for 
the negatively charged 
muon is more significant, increasing 
the cross-section approximately by $15-20 \%$. 
In Eq.(\ref{CROSS-SECTION}),
the sum goes over the available nuclear excitations, 
denoted by $\lambda$.
The nuclear structure effects are incorporated into  
${\cal M}_{\lambda 0}$, the bilinear combination of the nuclear matrix
elements between the ground state $|\tilde 0\rangle$  
and the excited states $\vert \lambda \rangle$ of the daughter 
nucleus. These are given by \cite{kub}:
\begin{equation}
\label{ME}
{\cal M}_{\lambda 0} = M_F |\langle \lambda | F | \tilde 0 \rangle |^2 +
 M_{GT} |\langle \lambda | GT | \tilde 0 \rangle |^2 +
 M'_{GT} \Lambda 
\end{equation}
The coefficients $M_i$ are obtained by the Foldy-Wouthuysen transformation 
 of the weak Hamiltonian where the 
terms up to third order in the momentum transfer
$q/M$ are kept ($M$ is the nucleon mass)\cite{kub}. 
The first matrix element squared  of 
Eq. (\ref{ME}) is:
\begin{equation}
\label{ME1} 
| \langle \lambda | F | \tilde 0 \rangle |^2 =
4 \pi \sum_{J} | \langle \lambda,J || 
t_{-} j_{J}(qr) Y_{J} || \tilde 0 \rangle|^2
\end{equation}
Here, $\vec{q}$ is the momentum transfer, $q=|\vec{q}|$,
$\vec{\sigma}$ and  $t_{-}$ refer to the 
nucleon spin Pauli matrices and isospin-lowering operator, respectively,
$||...||$ stands for the standard definition of the reduced 
matrix elements, $j_L$ are the spherical Bessel functions and $Y_J$ are the
spherical harmonics.
The remaining combinations of the matrix elements are:
\begin{equation}
\label{ME2}
| \langle \lambda | GT | \tilde 0 \rangle |^2 =
4 \pi \sum_{l,J} | \langle \lambda,J || 
t_{-} j_{l}(qr) 
\left[ Y_{l} \times \vec{\sigma} \right]_{J}
|| \tilde 0 \rangle|^2~,
\end{equation}
\begin{eqnarray}
\label{ME3}
\Lambda & = & 4 \pi \left( \frac{5}{6} \right)^{1/2} \sum_{l,l',J}
(-1)^{(l-l')/2+J} 
((2 l+1)(2l'+1))^{1/2}
 \nonumber \\ 
 & \times & \left\{ \matrix{
l& l'& 2 \nonumber \cr 0 & 1 & 2 \nonumber \cr }
\right\} 
  \left\{ \matrix{ 
1 & 1 & 2 \nonumber \cr l' & l & J \nonumber \cr } 
\right\}
 \langle \lambda,J|| 
t_{-}j_{l}(qr) \left[ Y_{l} \times \vec{\sigma} \right]_{J}
|| \tilde 0 \rangle 
 \nonumber \\ 
 & \times & \langle \lambda,J|| 
t_{-}j_{l'}(qr) \left[ Y_{l'} 
\times \vec{\sigma} \right]_{J}
|| \tilde 0 \rangle^{*}~.
\end{eqnarray}
Here, $[\quad \times \quad ]_J$ means the coupling to the total angular
momentum $J$. In our calculation we take into account states with $J 
\le 3$ with positive and negative parity. 

We perform the integration over angle $\theta$ in 
Eq.(\ref{CROSS-SECTION}) with the step 
size $\Delta \theta = 2 \pi /30$.
The single-particle matrix elements of the operators $O$
are calculated using the Hartree-Fock wave functions
in steps of $\Delta r$=0.1 fm in the radial coordinate. 

The $^{12}$C nucleus is not described well by a closed $p_{3/2}$  
subshell and configuration mixing is important in the ground state. 
Besides the RPA correlations,  
one of the important correlations is introduced by the pairing 
force. Here, we estimate the effect of pairing on the 
{\it exclusive and inclusive} neutrino cross-sections. 
In the expressions (\ref{CROSS-SECTION}-\ref{ME3}) for the cross-section,
two types of single-particle matrix elements enter:
a) those which do not contain the spin operator $\vec{\sigma}$
and
b) those which contain $\vec{\sigma}$. The two kinds have different
symmetry properties 
under time reversal. 
Consequently, the corresponding expressions for these
single-particle operators in terms of quasiparticles
are\cite{mot}: 
\begin{eqnarray}
\langle i^* | O \sigma |k^* \rangle
=
(u_i v_k - u_k v_i ) \langle i | O \sigma |k \rangle~,
\nonumber\\
\langle i^* | O  |k^* \rangle
=
(u_i v_k + u_k v_i ) \langle i | O  |k \rangle~,
\label{eqpair}
\end{eqnarray}
where the asterisks mean 
that the pairing has been taken into
account, $O$ is an operator that depends on $\vec{r}$ only, and the 
$u$ and $v$ are the coefficients of the Bogoliubov-Valatin 
transformation. 
In the case of $^{12}C$,
the $u$ and $v$ factors differ significantly  
from their values without pairing 
only for the $p_{3/2}$ and $p_{1/2}$ levels. 
This results in reduction factors: 
\begin{equation}
\zeta^2 = (u_{p_{1/2}} v_{p_{3/2}} -
v_{p_{1/2}} u_{p_{3/2}}
)^2~,
\end{equation}
which multiply the single-particle matrix elements of the operators
containing $\sigma$ in the expressions for the RPA 
matrix elements $\langle \lambda | O | \tilde 0 \rangle$. 
The evaluation of 
pairing effects in the cross-sections is done by introducing the 
factors $u_i v_k \pm u_k v_i$ 
in the matrix elements obtained from 
the RPA solutions without pairing. 

To obtain the cross-sections that allow for comparison with the
experimental data one has to fold the energy-dependent cross-section
of Eq.(\ref{CROSS-SECTION}) with 
a corresponding neutrino flux $f(E)$
\begin{equation}
\langle \sigma \rangle_f =
\int
dE \sigma (E) \tilde{f} (E)~,
\label{f1}
\end{equation}
where $\tilde{f}(E)$ is a properly normalized neutrino flux 
from an available neutrino source,
\begin{equation}
\tilde{f}(E) = \frac{f(E)}{\int_{E_0}^{\infty} dE' f(E')}~, 
\label{f2}
\end{equation}
and $f(E)$ is the initial (unnormalized) flux from the source.
Here, the value of $E_0$ depends on the neutrino source
 used in each experiment.
It is taken to be zero for the case of the electron neutrino 
\cite{kleinf}, 
while for the case of the muon neutrino source $E_0=E_{thr}$ \cite{alb},
where $E_{thr}$ is the threshold energy for the 
$^{A}Z (\nu_{\mu},\mu^-)$ reaction. 
We calculate the cross-sections with the neutrino energy 
steps $\Delta E = 1$MeV and $\Delta E = 5$MeV
for the electron and muon neutrino cases, respectively.
The spectra for the muon and electron neutrinos were 
taken from Refs.\cite{alb,kleinf}. The endpoint for the 
electron neutrino flux is $52.7$MeV, thus the cross-section 
for the ${^{A}Z}({\nu_{e}},e^-) $ reaction is sensitive to 
the low-energy excitations in the 
daughter nucleus, mainly the giant resonances.  
For the case of the muon neutrino,
the flux is cut at $E$= 260 MeV.

   The results of the calculations are shown in Table 1. 
We note first that the results obtained do not differ very much for the 
two interactions used and the cross-sections or capture rates agree 
typically within 
10$\%$ for the SGII and SIII interactions. 
In order to assess the influence of collective effects and 
of the RPA ground state correlations we have calculated each cross-section
or capture rate first in the Hartree-Fock approximation and then in the 
HF-RPA.  
The comparison shows that in the RPA the $\mu^-$-capture rates are reduced by 
30\% 
and 50\% in $^{12}$C and $^{16}$O, respectively. 
This is in agreement with the results in 
Ref.\cite{mek}.  
Note that in the present calculation of the $\mu^-$-capture rates
the SU4 assumption is not used and the vector, axial vector and 
induced pseudo-scalar contributions are directly calculated. 
  From Table 1 we see 
that the RPA effects are also very important in the ($\nu_e, e^-$) process.
In $^{12}$C and $^{16}$O the flux averaged ($\nu_e, e^-$) 
cross-section is reduced by 
nearly a factor two when one goes from HF to HF-RPA. 

One of the manifestations of the fact that $^{12}$C is not a closed 
shell is the need for 
a large suppression factor in order 
to reproduce the experimental exclusive cross-sections to the ground 
states of the A=12 nuclei\cite{wal,kol}. 
In Ref.\cite{kol} a reduction factor $\zeta^2$=3.88 - 4.13 was 
introduced  in the computation of 
exclusive processes, however these authors have not considered 
the influence of configuration mixing in the $^{12}$C ground state on 
the inclusive cross-sections to the excited states. In the present 
work, by including the pairing correlations we are able to treat the 
effects of such configuration mixing in both exclusive and inclusive 
cross-sections. The pairing calculations are done employing Eq. 
(\ref{eqpair}) 
and a $v_{p1/2}$=0.65, a choice in accordance with the value of the 
gap parameter $\Delta \simeq $3-4 MeV\cite{mot}. The 
pairing correction reduces the flux-averaged exclusive cross-sections 
and $\mu^-$-capture rates 
to the ground state of the daughter nucleus 
by factors 4-7 compared to 
the RPA results, depending on the type of reaction one calculates. The 
inclusive cross-sections to the excited states (denoted as $<\sigma^*>$) 
are affected less by 
the pairing correction but still the effect is sizable, reducing the 
$<\sigma^*>$ cross-sections by 25\%. The reduction due to pairing of the 
total inclusive processes is about a factor of 5 for the $(\nu_e, 
e^-)$ cross-section(because it is dominated by the transition to the 
$J^{\pi}=1^+$ ground state) and about 33\% reduction for the 
$(\nu_{\mu}, \mu^-)$ cross-section with respect to the RPA result. 
Note that the RPA and pairing correction decrease the 
cross-sections calculated in the HF approximation by a factor of 6-7 for 
$(\nu_e, e^-)$ and a factor of 2 for the $(\nu_{\mu}, \mu^-)$ and 
$\mu^-$-capture processes. These reductions are essential in achieving 
agreement with experiment. We emphasize that the agreement is 
achieved in all the quantities calculated by 
using the same value of the parameter $v_{p1/2}$. We did not 
attempt to find the best value for this parameter. In parenthesis of 
column 4 we show the results obtained when we use 
$v_{p1/2}$=0.60~. We should remark that the results in Table 1 
corresponding to interaction SGII are obtained with a slightly 
changed spin-orbit parameter (compared to the original SGII force) so 
as to reproduce in the RPA the experimental threshold energy. 
Our results for the $(\nu_{\mu}, \mu^-)$ inclusive 
cross-section of 13.5x10$^{-40}$cm$^{2}$ (for SGII) and 14.5x10$^{-40}$cm$^{2}$ 
(for SIII) should be compared with the recently revised experimental 
value of (11.2$\pm$0.3$\pm$1.8)x10$^{-40}$cm$^{2}$ from the LSND 
experiment\cite{kim}.

In $^{16}$O the nucleons form a good closed shell and pairing effects 
are not important. 
Except for the $\mu^-$-capture there are no experimental neutrino 
data. Our HF-RPA predictions in $^{16}$O for the flux-averaged 
inclusive $(\nu_{\mu}, \mu^-)$ cross-sections are 27.8x10$^{-40} cm^2$ 
(SGII) and 27.1x10$^{-40} cm^2$ (SIII) whereas for $(\nu_e, e^-)$ they 
are 17.2x10$^{-42} cm^2$ (SGII) and 16.9x10$^{-42} cm^2$ (SIII). The 
calculated HF-RPA $\mu^-$-capture rates are 1.05x10$^{5} s^{-1}$ (SGII) 
and 1.00x10$^{5} s^{-1}$ (SIII) to be compared with the experimental 
value of 0.98x10$^{5} s^{-1}$\cite{eck}.

In summary, we have shown that the use of the consistent HF-RPA 
scheme and the introduction of pairing corrections can successfully 
reproduce the experimental $\mu^-$-capture rates and neutrino 
cross-sections in $^{12}$C. Although the present approach is not as 
elaborate as an extended shell-model calculation or a quasi-particle 
RPA it contains the basic features that emerge from such more 
extensive methods.  

We wish to thank A. Hayes and W.C. Louis for discussions. 
Two of us (N.A. and O.K.V.) 
thank X. Campi and D. Vautherin for their hospitality 
at the Division de Physique Th\'eorique in Orsay. 
DPT of IPN-Orsay is a Unit\'e de Recherche 
des Universit\'es Paris XI et Paris VI associ\'ee au CNRS.

\begin{table}
\caption{
Flux-averaged cross-sections and $\mu^-$-capture rates in $^{12}$C. 
$\sigma_{exc}$ stands for exclusive cross-sections to the $^{12}$N 
ground state, $\sigma^{*}$ is the inclusive cross-section to excited 
states. The capture rates $\omega_{g.s.}$ and $\omega_{tot}$ correspond to the 
partial rate to the ground state and to the total rate, respectively. 
The $(\nu_{\mu},\mu^-)$ cross-sections are in $10^{-40}cm^2$, 
$(\nu_{e},e^-)$ cross-sections are in $10^{-42}cm^2$, capture rates 
are in $10^{4}s^{-1}$. For each quantity, values calculated with SGII 
and SIII are in the upper and lower row, respectively. 
The results in brackets correspond to the choice 
$v_{p1/2}$=0.60~. 
}

\vspace{3mm}
\begin{tabular}{|c|c|c|c|c|c|}
  Channel    & HF & HF & HF- & HF-RPA & Exp. \\
     &    &+pair. & RPA    & +pair. &     \\
\hline
$(\nu_{\mu}, \mu^-)$ & & & & & \\
  $\sigma_{exc}$ & 4.28 & 0.61 & 3.35 & 0.39 (0.64) & 
0.66$\pm$0.1$\pm$0.1\cite{kim} \\
  & 4.70 & 0.67 & 3.80 & 0.50 (0.77) & \\
  & & & & & \\ 
  $\sigma^{*}$ & 22.7 & 16.1 & 17.7 & 13.1 
(13.7) &  \\
  & 24.1 & 18.0 & 18.6 & 14.0 (14.4) & \\
  & & & & & \\
 $\sigma_{inc}$  & 27.0 & 17.7 & 21.1 & 13.5 (14.3)  & 
11.2$\pm$0.3$\pm$1.8\cite{kim} \\
  & 28.8 & 18.7 & 22.4 & 14.5 (15.2) & \\
\hline
 $(\nu_{e}, e^-)$ & & & & & \\
  $\sigma_{exc}$ & 78.1 & 10.0 & 54.8 & 7.1 (11.4) & 
10.5$\pm$1.$\pm$1.\cite{allen,kraka}  \\
  & 100.4 & 14.2 & 68.2 & 10.1 (16.0) & 
8.2$\pm$0.65$\pm$0.75\cite{kleinf}\\
  & & & & & \\ 
  $\sigma^{*}$ & 8.6 & 8.8 & 8.3 & 5.6 
(6.2) & 
 6.4$\pm$1.45$\pm$1.4\cite{zei}
 \\
  & 14.0 & 10.7 & 8.1 & 6.4 (6.7) & 3.6$\pm$2.7\cite{kraka} \\
  & & & & & 5.7$\pm$0.6$\pm$0.6\cite{ath97} \\ 
  & & & & & 8.6$\pm$1.2$\pm$1.5\cite{kleinf} \\ 
  & & & & & \\ 
 $\sigma_{inc}$  & 90.6 & 18.8 & 63.2 & 12.9 (17.6)  & 
 14.1$\pm$2.3\cite{kraka} \\ 
  & 114.4 & 24.9 & 76.3 & 16.5 (22.7) & 16.8$\pm$1.7\cite{kleinf} \\ 
\hline
$\mu^-$-capt. & & & & & \\
 $\omega_{g.s.}$ & 3.61 & 0.51 & 3.24 & 0.41 (0.67) & 
0.60$\pm$0.04\cite{miller,giffon} \\ 
  & 3.70  & 0.53  & 3.48 & 0.45 (0.73) & \\ 
  & & & & & \\ 
   $\omega_{tot}$ & 8.0 & 3.87 & 6.87 & 3.09 (3.48) & 
3.7$\pm$0.1\cite{eck} \\ 
  & 8.4  & 4.10  & 7.22 & 3.23 (3.64) & \\
\end{tabular}
\end{table}


\begin{thebibliography}{99}
\bibitem{bah} J.N. Bahcall, {\it Neutrino Astrophysics} (Cambridge 
University Press, New York, 1989); J.N. Bahcall and M.H. Pinsonneault, 
Revs. Mod. Phys. {\bf 64}, 885 (1992).
\bibitem{wal}
J.D.Walecka, in {\it Muon Physics}, 
vol. II, ed. by V.M.Hughes and C.S.Wu (Academic Press, New York, 1975); H. 
\"Uberall and B.A. Lamers, Phys. Rev. C {\bf 6}, 1911 (1972); J.S. 
O'Connell, T.W. Donnelly and J.D. Walecka, Phys. Rev. C {\bf 6}, 719 
(1972).  
\bibitem{kub}
T. Kuramoto, M. Fukugita, Y. Kohyama and K. Kubodera, Nucl.Phys. {\bf 
A512}, 711 (1990).
\bibitem{kol} E. Kolbe, K. Langanke and S. Krewald, Phys. Rev. C {\bf 
49}, 1122 (1994). 
\bibitem{ath96} C. Athanassopoulos {\it et al}., Phys. Rev. Lett. {\bf 77}, 
3082 (1996).
\bibitem{alb} M. Albert {\it et al}., Phys. Rev. C {\bf 51}, R1065 (1995).
\bibitem{boh} O. Bohigas, A.M. Lane and J. Martorell, Phys. Rep. 
{\bf 51}, 267 (1979).
\bibitem{ber} G.F. Bertsch and S.F. Tsai. Phys. Reports {\bf 18C}, 126 
(1975).
\bibitem{mek} N. Van Giai, N. Auerbach and A.Z. Mekjian, Phys. Rev. Lett. 
{\bf{46}},1444 (1981); N. Auerbach and A. Klein, Nucl. Phys. {\bf 
A422}, 480 (1984).
\bibitem{bei75} M. Beiner {\it et al}., Nucl. Phys. {\bf A238}, 29 (1975).
\bibitem{gia81} N. Van Giai and H. Sagawa, Phys. Lett. B {\bf 106}, 
215 (1981).
\bibitem{aue83} N. Auerbach and A. Klein, Nucl. Phys. {\bf A375}, 77 
(1983). 
\bibitem{col} G. Col\`o, N. Van Giai, P.F. Bortignon and R.A. Broglia, 
Phys. Rev. C {\bf 50}, 1496 (1994).
\bibitem{mot} A. Bohr and B.R. Mottelson, {\it Nuclear Structure} 
(Benjamin, New York, 1969).
\bibitem{kleinf} J. Kleinfeller {\it et al}., in {\it Proceedings of 
the PANIC 93 Conference}, Perugia, Italy (1993).
\bibitem{kim} 
C. Athanassopoulos {\it et al}., submitted to Phys. Rev. {\bf C}.
\bibitem{eck} M. Eckhause, R.T. Siegel, R.E. Welsh and T.A. Filippas, 
Nucl. Phys. {\bf 81}, 575 (1966).
\bibitem{allen} R.C. Allen {\it et al}., Phys. Rev. Lett. {\bf 64}, 1871 
(1990).
\bibitem{kraka} D.A. Krakauer {\it et al}., Phys. Rev. C {\bf 45}, 2450 
(1992).
\bibitem{zei} B. Zeitnitz, KARMEN collaboration, Prog. Part. Nucl. 
Phys. {\bf 32}, 351 (1994).
\bibitem{ath97} C. Athanassopoulos {\it et al}., Phys. Rev. C {\bf 
55}, 2078 (1997).
\bibitem{miller} G.H. Miller {\it et al}., Phys. Lett. {\bf 41B}, 50 
(1972).
\bibitem{giffon} Giffon {\it et al}., Phys. Rev. C {\bf 21}, 241 (1981).
\end{thebibliography}
\end{document}